\documentclass[prb,twocolumn,aps]{revtex4-2}
\usepackage{graphicx,amsmath,xcolor}
\usepackage{hyperref}
\usepackage{subcaption}
\usepackage{float}
\usepackage{svg}
\usepackage{silence}
\usepackage{mathtools}
\usepackage[T1]{fontenc}
\DeclarePairedDelimiter{\ket}{\lvert}{\rangle}%
\DeclarePairedDelimiterX\innerp[2]{\langle}{\rangle}{#1\delimsize\vert\mathopen{}#2}%
\DeclarePairedDelimiterX\braket[2]{\langle}{\rangle}{#1\delimsize\vert\mathopen{}#2}%
\DeclarePairedDelimiterX\braketOP[3]{\langle}{\rangle}{#1\,\delimsize\vert\,\mathopen{}#2\,\delimsize\vert\,\mathopen{}#3}%
\DeclarePairedDelimiterX\ketbra[2]{\lvert}{\rvert}{#1\delimsize\rangle\!\delimsize\langle#2}%
\DeclarePairedDelimiterX\outerp[2]{\lvert}{\rvert}{#1\delimsize\rangle\!\delimsize\langle#2}%
\DeclarePairedDelimiterX\projector[1]{\lvert}{\rvert}{#1\delimsize\rangle\!\delimsize\langle#1}%

\begin{document}

    \title{Assessing quantum dot SWAP gate fidelity using tensor network methods}
    \author{Jacob R. Taylor}
    \author{Nathan L. Foulk}
    \author{Sankar Das Sarma}
    \affiliation{Condensed Matter Theory Center and Joint Quantum Institute, Department of Physics, University of Maryland, College Park, Maryland 20742-4111 USA}
\begin{abstract}
    Advanced tensor network numerical methods are used to explore the fidelity of repeated SWAP operations on a system comprising 20 to 100 quantum dot spin qubits in the presence of valley leakage and electrostatic crosstalk. The fidelity of SWAP gates is largely unaffected by Zeeman splitting and valley splitting, except when these parameters come into resonance. The fidelity remains independent of the overall valley phase for valley eigenstates, while for generic valley states, some minor corrections arise. We analyze the fidelity scaling for long qubit chains without valley effects, where crosstalk represents the only error source.
\end{abstract}
\maketitle 

\section{Introduction}
A standard operation in quantum computing is the SWAP gate, which allows for the exchange (or ``swapping'') of the quantum states between two qubits.
The SWAP gate has important applications in quantum computing, such as in quantum error correction~\cite{gottesman1997stabilizer}, measurement schemes~\cite{barenco1997stabilization}, and quantum state engineering~\cite{nielsen2010quantum}.
The root $\sqrt{\text{SWAP}}$ is entangling and thus, in combination with arbitrary single-qubit gates, allows for the implementation of general unitary operations, sufficient for universal quantum computation~\cite{loss1998quantum}.
This fact, combined with the SWAP gate's ubiquity within quantum error correction, makes the creation of high fidelity SWAP gates essential for building a quantum computer on any platform. 

Si-based quantum dot spin qubits have emerged as a promising candidate for realizing quantum computers due to their long coherence times~\cite{kobayashi2021engineering,Burkard2023}.
Both $^{28}$Si and $^{30}$Si are common spin-0 isotopes, and thus it is possible to remove the decoherence arising from nuclear spin noise by isotopic purification~\cite{Witzel2010}.
In addition, electrical gate operations implemented directly from the Heisenberg interaction between different qubits, such as SWAP gates, have durations on the order of 1 ns~\cite{medford2013quantum}.
These long coherence times arising from such isotopic purification, combined with the available short gate operation times, make Si especially well-suited for hosting qubits.
Most importantly, silicon-based qubits can be easily integrated into the existing semiconductor industry, allowing for more straightforward scalability. 
A silicon-based quantum computer platform could potentially host millions of qubits in a small chip similar to existing CMOS-based integrated circuits, and there has been spectacular recent progress in producing scalable multiqubit Si circuits, with the largest devices---currently fabricated in commercial foundries---now at twelve qubits~\cite{neyens2023,IntelSite}. 
Based on this trajectory, it seems that devices of twenty or more dots are in the not so distant future. 
It is therefore both timely and important to consider SWAP gates in large spin qubit systems.
Standard theoretical and computational techniques, such as exact diagonalization, cannot handle the simulation of large numbers of qubits since the Hilbert space size grows exponentially with the number of qubits.  
The goal of the current work is to introduce the powerful tensor network techniques to study systems of many entangled spin qubits in order to establish the tensor network method as a viable theoretical tool in the modeling of spin qubit arrays.

Tensor networks provide a framework for representing many-body quantum states and operators in a computationally compact and efficient way, allowing for accurate simulations of large quantum systems which are intractable with direct methods~\cite{biamonte2017tensor}.
Acting on matrix product states (MPS) with tensor network time evolution methods such as TEBD~\cite{daley2004time} or TDVP~\cite{haegeman2011time} provides a technique to accurately approximate the time evolution of interacting quantum systems and can be used to investigate model-intrinsic sources of error.
Systems of hundreds of qubits that would be utterly intractable through direct simulation can often be represented efficiently by restricting one's system to an approximate low-entanglement subspace using tensor networks.
This is what we do in the current work by studying spin qubit arrays up to 100 qubits using the tensor network technique, establishing the feasibility of tensor networks as an effective tool for spin qubit theories and analyses.

The valley degeneracy in Si introduces additional states to the qubit manifold and possible avenues for leakage and relaxation~\cite{Koiller2001,Friesen2010,Zwanenburg2013,Burkard2023}.
Here we seek to expand previous work into the effects of valley states on spin qubit devices by directly including the additional valley states within our simulation and quantifying the effects of valley leakage.
We also include the effects of crosstalk in the theory as explained later in the text.
Previous work into the fidelity of sequences of SWAP gates on a spin qubit chain has looked into the effects of charge noise and dissipation on the fidelity of SWAP gates~\cite{foulk2022dissipation,throckmorton2020fidelity}. 
Such work did not include dynamical valley states or their interactions and was done only on small-scale systems due to computational constraints in exact diagonalization. 
Our work can address tens to hundreds of spin qubits in contrast to all earlier works on the subject by using state-of-the-art methods within tensor networks to accurately model the effect of the initial valley state and different experimentally relevant parameters on the fidelity of a sequence of chained SWAP gates.

We first introduce our spin qubit model and describe the tensor network based numerical methods used to perform the calculations. 
We then present the results of those numerical calculations, demonstrating the effect of valley splitting, Zeeman splitting, and SWAP exchange strength on the fidelity of SWAP operations. 
Of particular interest are the effects of the spin-valley coupling and its phase on different initial valley states. 
We conclude with an investigation into the single gate fidelity scaling up to 100 spin qubits, isolating the influence of crosstalk.


\section{Model}

In exchange-coupled spin qubits~\cite{loss1998quantum}, the spin states of individual electrons serve as the computational basis.
The basis states of our model are labeled by the valley and spin degrees of freedom. 
The Hilbert space of each electron includes the lowest two valley states  $\ket{\pm}$ of each dot, corresponding to $k=\pm z$, along with, within each valley, the spin states $\ket{\uparrow}$ and $\ket{\downarrow}$.
It is reasonable to focus on the spin-valley mixing and not consider orbital and valley effects separately since the states $\ket{\pm}$ are, in actuality, valley-orbit states. 
This is because of large valley-orbit couplings in Si/SiGe quantum dots which make the valley index in general not a good quantum number~\cite{Friesen2010}. 
Nonetheless, the energy gap between the two lowest energy eigenstates is always referred to as the ``valley splitting''.
Our work focuses on the effects of spin-valley mixing specifically and the associated leakage. 

Our model is a one-dimensional (1D) spin chain with both valley and spin degree of freedom using the following Hamiltonian:
\begin{multline}
H=\sum_{n=1}^{L-1} J_n \left(\boldsymbol{\sigma}_n\cdot \boldsymbol{\sigma}_{n+1}+1\right)\left(\boldsymbol{\tau}_n\cdot \boldsymbol{\tau}_{n+1}+1\right)+\\ h \sum_n^L \sigma_n^{z}+ \Delta \sum_n^L \tau_n^z+\\
\frac{\gamma_1}{2}\sum_n^L\left(\tau_n^x\sigma_n^x+\tau_n^y\sigma_n^y\right)+\frac{\gamma_2}{2}\sum_n^L \left(\tau_n^y\sigma_n^x-\tau_n^x\sigma_n^y\right)
\label{hamiltonian}
\end{multline}

where $L$ is the number of qubits in the spin chain, $h$ is the spin Zeeman splitting, $\Delta$ is the valley splitting and $\gamma_1$,$\gamma_2$ are the real and imaginary parts of the spin-valley coupling $\gamma=\gamma_1+i \gamma_2$.
$\boldsymbol{\sigma}_n=(\sigma_n^x,\sigma_n^y,\sigma_n^z)$ is the $n^\text{th}$ site Pauli vector in the spin basis, while $\boldsymbol{\tau}_n$ is the same but in the valley basis. The Hamiltonian defined by Eq.~\ref{hamiltonian} should be thought of as the minimal model for a spin qubit array including spin-valley coupling, which is sufficient for our purpose of establishing the power of the tensor network techniques in studying systems of spin qubits.  Experimental spin qubit systems are likely to have additional nonessential complications, which would vary from sample to sample, which cannot be captured in a universal theoretical description.  These sample-dependent complications must necessarily focus on specific samples, thus losing the generality and the universality of the minimal theory which must be developed first.

The role of the spin-valley coupling becomes clear in the matrix representation of the single-site Hamiltonian,
\begin{equation}
    H_n = \begin{pmatrix} 
            h + \Delta & 0 & 0 & 0 \\
            0 & h - \Delta & \gamma & 0 \\
            0 & \gamma* & \Delta - h & 0 \\
            0 & 0 & 0 & -\Delta - h 
    \end{pmatrix}.
\end{equation}
The valley splitting is material-specific and cannot be controlled. 
In the case of zero magnetic field, the two lowest states are the $\ket{- s}$, where $s = \uparrow$ or $\downarrow$.
As the magnetic field strength increases, the electron forms a spin qubit two-level system, but this becomes entangled by spin-valley mixing as $h \to \Delta$.
But this spin-valley mixing is only relevant between $\ket{- \uparrow}$ and $\ket{+ \downarrow}$ since the gap between these states shrinks as the magnetic field is increased.
The gap between the $\ket{-\downarrow}$ and $\ket{+\uparrow}$ is at a minimum at zero magnetic field and only grows in the spin qubit regime ($h>0$). 
Therefore, the coupling between these states is negligible.
The main concern of this work is the effect of spin-valley leakage, which is the loss of spin fidelity as a result of coherent evolution into a nearby valley state of a different spin, as the spin and valley eigenstates get entangled by the last two terms in Eq.~\ref{hamiltonian}.
Therefore, the other couplings can be set to zero, as transitions between different valley states of the same spin do not impact our computational state, and intra-valley spin relaxation is beyond the scope of this work (and is very weak in Si anyway).

The spin-valley coupling originates in both Rashba and Dresselhaus spin-orbit coupling of the host material~\cite{Tahan2014}. When the external magnetic field is aligned in the [110] direction, the relevant spin-valley matrix element simplifies~\cite{huang2014spin} to 
\begin{equation}
    \gamma = \frac{2 m^* (\alpha_R + \alpha_D) r^{-+}_{x}}{\hbar}\Delta,
\end{equation}
where $r^{-+}_x$ is the electric dipole matrix element between valley states, $m^*$ is the effective electron mass, and $\alpha_R$ and $\alpha_D$ are the Rashba and Dresselhaus spin-orbit interaction constants, respectively. Reasonable estimates of these parameters for silicon quantum dots lead to an order-of-magnitude estimate of $\frac{|\gamma|}{\Delta} \sim 1/500$.
The precise number for this coupling is not important for our considerations in this work, where the goal is to establish the efficacy of the tensor network techniques.

The SWAP operation is performed using a $\frac{\pi}{4}$ pulse Heisenberg interaction between the swapping sites. 
Ideally, the exchange coefficient should be zero for non-swapping sites. 
In reality, this is never true~\cite{Mills2022,Noiri2022,Takeda2022,Philips2022}. 
The exchange coefficient $J_n$ is set as follows:

\begin{equation}
J_n = 
\left\{
    \begin{array}{lr}
        J_0, & \text{if } n \neq l\\
        J_{\text{SWAP}}, & \text{if } n = l
    \end{array}
\right\},
\end{equation}

where the SWAP gate is between sites $l$ and $l+1$.
For simplicity, $J_0$, $\gamma$, $\Delta$ and $h$ are all taken to be site independent. 
These nonessential assumptions are easy to relax in our method; they have little impact on the simulation runtimes.
Since these parameters would vary among experimental samples, it makes little sense to include them as site-dependent in our work although it is straightforward to do so at the cost of having a large number of unknown qubit-dependent parameters in the theory.

The initial state is a product state between an initial spin state $\ket{\psi_i}_s$ and the valley state $\ket{\psi_i}_v$ up to a normalization factor,
\begin{equation}
\ket{\psi_i}_v \propto (1-\alpha)\ket{-...-}+\alpha\ket{-_x...-_x},
\label{eqn:initialstate}
\end{equation}
where
$$\ket{-_x}=\frac{1}{\sqrt{2}}\left(\ket{+}-\ket{-}\right).$$
When $\alpha>0$, this simply represents the deviation from the ideal case, which is when each electron is initialized in the valley down state.

The total fidelity for our sequence of SWAP gates is defined as:
\begin{equation}
F_{\text{tot}}=\text{Tr}_s\left[\text{Tr}_v[U\rho_i U^\dagger]\text{Tr}_v[R \rho_i R^\dagger]\right],
\label{eqn:fidelitymeasure}
\end{equation}
 where $R$ is an operator which performs the SWAP gate sequence with perfect fidelity, $U$ represents the actual SWAP gate
sequence with errors, and $\rho_i=\ketbra{\psi_i}{\psi_i}$ is the initial state of the system.
$\text{Tr}_s[...]$ and $\text{Tr}_v[...]$ are the partial trace operators over spin and valley degrees of freedom, respectively.
In our case, the SWAP sequence transports a spin state from one side of the spin chain to the other.
The transport is performed by swapping sites $1\leftrightarrow2$, then $2\leftrightarrow3$, and so forth until the first spin state is transported to the end of the chain.
The effective single gate fidelity $F=(F_{tot})^{1/(L-1)}$ is used to be able to compare gate sequences of different lengths.

Our absolute fidelities are not relevant from an experimental perspective. 
The calculations here obviously cannot be used to predict fidelities of actual physical devices, since those values will depend on a plethora of unknown variables that are unique to each experimental setup, not to mention the many error mechanisms that are not included in our simulations.
The calculated fidelities are only useful in a relative sense and can only give information about the general model of interacting spin qubits, which is universal to all spin qubit samples.
If system dependent parameters are available, our technique can easily be used to predict the behavior of individual samples.
For example, one can compare the infidelities of our model from spin-valley mixing relative to the infidelities caused by residual exchange.
But the exact value of infidelity for a single calculation has limited experimental application. 

Since each physical spin qubit consists of spin and valley states, each can be split into two separate two-level sites in the tensor network representation.
The two-level tensors are arranged in two rows, with the top row representing the spin states and the bottom row representing the valley states.
To construct the tensor network Hilbert space, the sites are ordered such that the $2j-1$ and $2j$ MPS sites map to the $j$th physical qubit's spin and valley states, respectively.
The coupling between those sites $2j-1$ and $2j$ corresponds to the spin-valley coupling $\gamma$.
The interweaving of the two states is ideal because it minimizes the number of long-range interactions necessary within the Hamiltonian, thus improving the efficiency of the MPS algorithm.

\begin{figure}[H]
    \centering
    \includegraphics[width=0.40\textwidth]{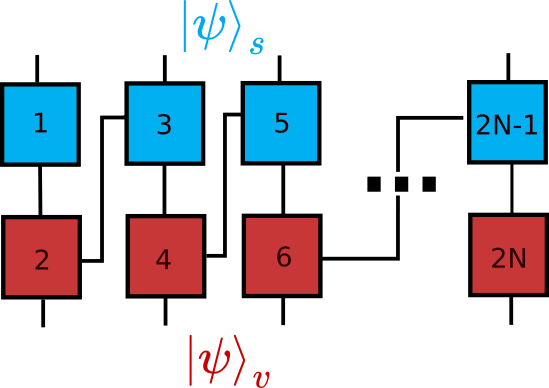}
     \caption{Tensor network diagram for the MPS of the spin chain.
        The blue and red sites represent the spin ($\ket{\psi_s}$) and valley ($\ket{\psi_v}$) degrees of freedom respectively.
        The tensors are ordered so they snake between the spin and valley degrees of freedom to minimize long-range interactions.}
     \label{fig:TensorDiagram}
\end{figure}

The system is initialized by constructing the spin state MPS and the valley state MPS independently and then interspersing them with standard tensor network operations to build a single total MPS for the entire system.
In the case of the random spin state MPS, an initial MPS of bond dimension $M=10$ with complex matrix elements is generated.
The time evolution of the total MPS is performed using Time Evolving Block Decimation (TEBD)~\cite{daley2004time} with SWAP gate time $T=\frac{\pi/4}{J_\text{SWAP}}$.
The state is mapped to a rotating reference frame to correct for the background rotation caused by the external magnetic fields.
This simple mapping is achieved by applying a set of local unitary operations to all sites $U_r=\Pi_{n=1}^L \exp(i h \sigma^z_n T)$ when computing the fidelity.
The resulting MPS is then converted into a projector matrix product operator (MPO), used in Eq.~\ref{eqn:fidelitymeasure}. 
For each of our results using tensor network methods, the calculations were repeated for smaller chain lengths ($L<15$), and independently confirmed using exact diagonalization. 
There was no significant discrepancy between calculated fidelities, both in a qualitative and quantitative sense. 
This gives us confidence that our calculations for longer chain lengths are well-grounded in reality and provides a clear success story for the use of tensor networks in spin qubit systems.

\section{Calculations}
As discussed before, our gate sequence consists of interleaved SWAP gates on a Heisenberg-coupled spin chain. 
The sequence swaps the first and second spins, then the second and third spins, iterating through the entire chain.

\begin{figure}[H]
    \centering
    \includegraphics[width=0.45\textwidth]{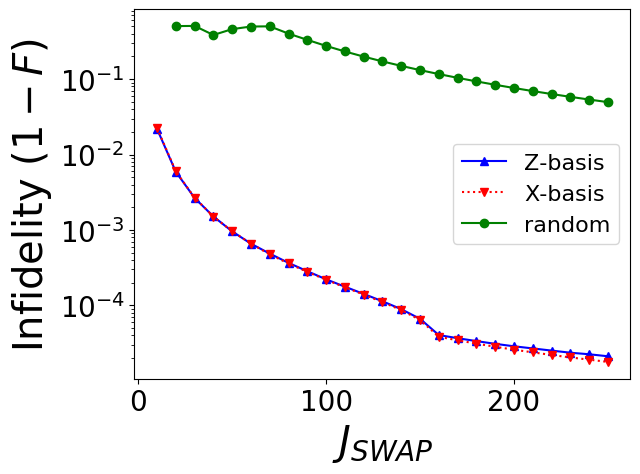}
    \caption{Single gate infidelity $(1-F)$ of a $L=50$ spin chain initialized with $\ket{\uparrow_x\downarrow_x\downarrow_x...\downarrow_x}$, $\ket{\uparrow\downarrow\downarrow...\downarrow}$ and a random spin state for $\alpha=0$, $h=750$, $\Delta=500$, $\gamma_1=1$ and $\gamma_2=0$.}
    \label{fig:JSwapFidelity}
\end{figure}

To assess the effects of different parameters on the fidelity of SWAP gates, it is reasonable to first consider the impact of the SWAP gate exchange coupling $J_{\text{SWAP}}$.
The larger $J_{\text{SWAP}}$ is relative to the residual exchange $J_0$, the larger the fidelities of the SWAP gate will be since this would trivially make the SWAP gates faster. 
Our results, as illustrated in Fig.~\ref{fig:JSwapFidelity}, affirm this expectation. 
This observation is consistent across different initial spin basis states, underscoring the robustness of the relationship.

It is also apparent that the choice of initial spin state is crucial. Being able to represent and manipulate entangled states is a critical quality of a successful quantum computer, and in this case, our results suggest that anything short of a product state leads to dramatically higher infidelities. It is worth noting that the only sources of error present in this calculation are crosstalk and spin-valley coupling. 
Crosstalk arises from the unintended coupling among different spin qubits because of the electrostatic control used in  gate operations~\cite{Buterakos2018,throckmorton2020fidelity}.
Even increasing $J_{\text{SWAP}}$ to relatively large values cannot compensate for the errors resulting from crosstalk and valley degeneracy.

It is worth noting that in all results, the relevant parameters $\Delta$, $h$, $J_{\text{SWAP}}$, $\gamma_1$ and $\gamma_2$ are in units of $J_0$, the residual exchange. 
In a recent experiment, the residual exchange was measured on the average to be about $J_0\approx 40$ kHz~\cite{Philips2022}.

\begin{figure*}
    \centering
     \begin{subfigure}{0.35\textwidth}
         \centering
         \includegraphics[width=\textwidth]{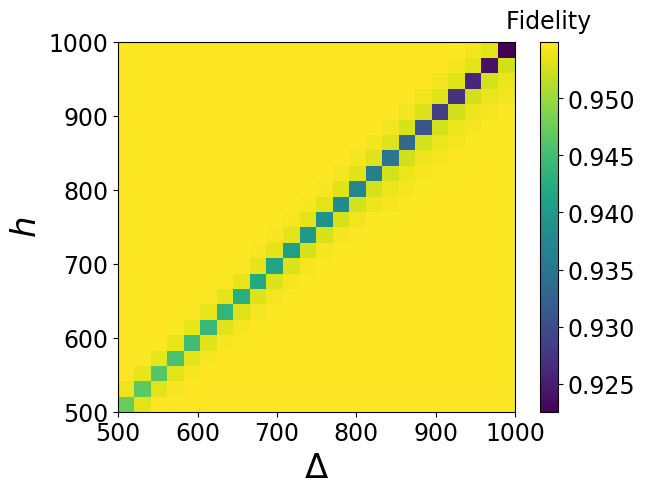}
         \caption{}
         \label{fig:DvHa}
     \end{subfigure}
     \begin{subfigure}{0.35\textwidth}
         \centering
         \includegraphics[width=\textwidth]{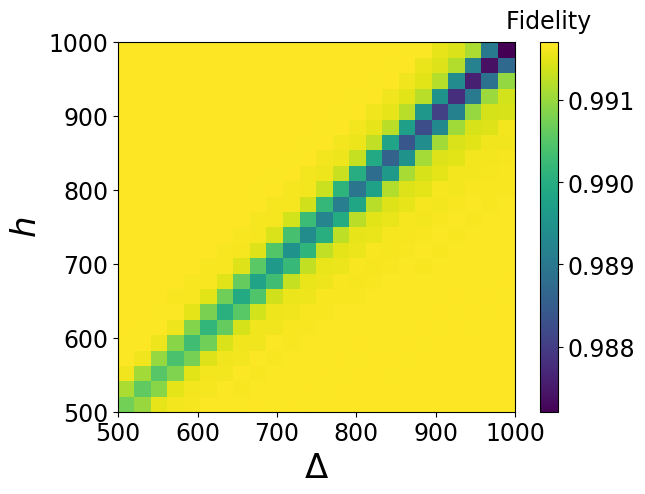}
         \caption{}
         \label{fig:DvHb}
     \end{subfigure}
     \begin{subfigure}{0.35\textwidth}
         \centering
         \includegraphics[width=\textwidth]{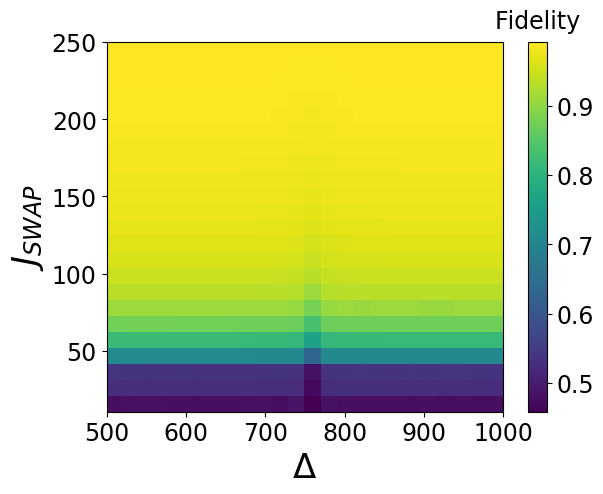}
         \caption{}
     \end{subfigure}
     \begin{subfigure}{0.35\textwidth}
         \centering
         \includegraphics[width=\textwidth]{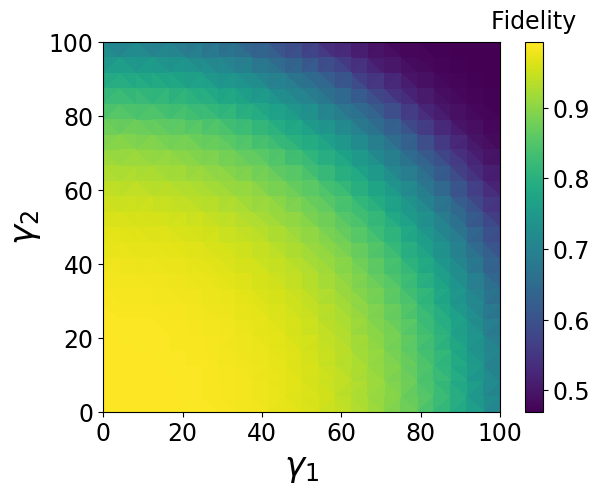}
         \caption{}
     \end{subfigure}
     \caption{Single gate fidelity for variable $h$ and $\Delta$ for a $L=20$, $\alpha=0$, random spin initial state with parameters $\gamma_1=\Delta/500$, $\gamma_2=0$, (a) $J_{\text{SWAP}}=100$ and (b) $J_{\text{SWAP}}=250$. (c) Single gate fidelity for variable $J_{\text{SWAP}}$ and $\Delta$ of a $L=20$, random spin, $\alpha=0$ initial state.
     The other parameters used were $h=750$, $\gamma_1=\Delta/500$, $\gamma_2=0$.
     The small deviation at $h\approx \Delta$ can be seen when $\Delta$ and $h$ are in resonance. (d) Single gate fidelity for variable $\gamma_1$ and $\gamma_2$ with $h=750$ and $\Delta=500$, and a  $L=20$, $\alpha=0$, random initial spin state.}
     \label{fig:DvH}
\end{figure*}

We next consider the effects of the spin splitting $h$, valley splitting $\Delta$, and spin-valley coupling $\gamma = \gamma_1 + i \gamma_2$. In Fig.~\ref{fig:DvH}, there is a significant impact on the fidelity of the SWAP sequence when the spin and valley splittings come into resonance ($h\approx \Delta$).
Large values of $J_{\text{SWAP}}$ broaden the $h \approx \Delta$ resonance regime, though they also suppress the infidelity peak in that regime. For example, in Fig.~\ref{fig:DvHa}, the error attributable to spin-valley coupling makes up about 40\% of the total error. However, for a larger value of $ J_\text{SWAP}$ in Fig.~\ref{fig:DvHb}, the spin-valley mixing makes up about 30\% of the error.

The magnitude $|\gamma|=\sqrt{\gamma_1^2+\gamma_2^2}$ of the spin valley coupling significantly impacts the fidelity of the SWAP gate.
A larger $\gamma$ is detrimental to the SWAP gates' reliability due to local spin-valley state mixing.
This can be seen in Fig.~\ref{fig:DvH}, and our results suggest that the magnitude of $\gamma$ dominates over its phase in the case of repeated SWAP gates.
This is a non-obvious finding of our calculations.

\begin{figure*}
    \centering
    \begin{subfigure}{0.4\textwidth}
        \centering
        \includegraphics[width=\columnwidth]{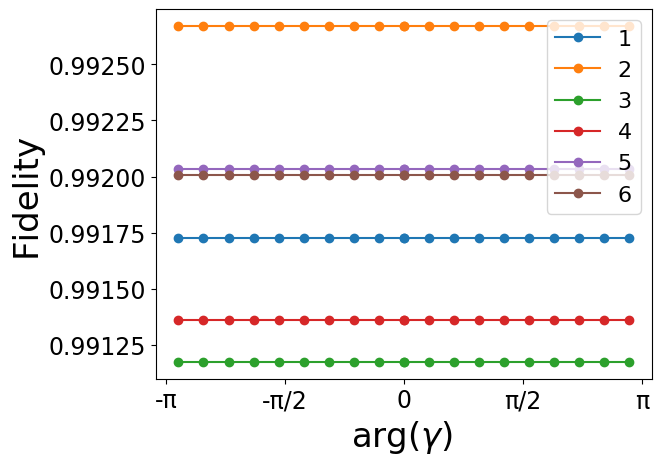}
        \caption{}
    \end{subfigure}
    \begin{subfigure}{0.4\textwidth}
        \centering
        \includegraphics[width=\columnwidth]{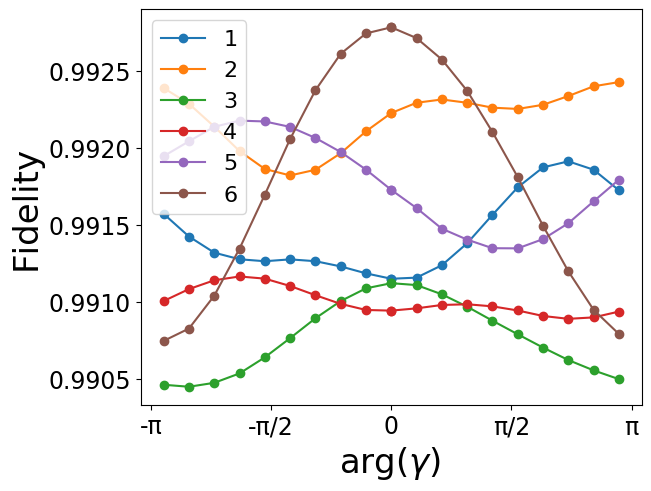}
        \caption{}
    \end{subfigure}
    \caption{        
	Single gate fidelity for six random $L=20$ initial spin states.
        This was calculated for different phases of $\gamma=10e^{i\theta}$.
        The initial states had (a) $\alpha=0$ (b) $\alpha=1$.
        In both cases $h=30$ and $\Delta=100$.
    }
    \label{fig:alpha0_and_alpha1}
\end{figure*}

The implication that the valley phase is unimportant for spin qubit SWAPs is surprising and important for experimental systems.
To highlight this phenomenon, the fidelities of the SWAPs are examined as a function of only the spin valley coupling phase $\theta=\text{arg}(\gamma)$. 
The SWAP fidelity is shown for six random initial spin states at $\alpha=0$ and $\alpha=1$ in Fig.~\ref{fig:alpha0_and_alpha1}.
For the $\ket{-\ldots-}$ initial valley state ($\alpha = 0$), there is absolutely no valley phase dependence on fidelity, and only $|\gamma|$ has any effect.
The irrelevance of the valley phase is confirmed numerically up to machine precision. 
Analytically, if either of the spin or valley states is initialized to a $z$ basis state, the effect of $\text{arg}(\gamma)$ reduces to a global phase, which does not enter into the fidelity.

However, there is an obvious phase dependence when the valley state is initially $\ket{-_x\ldots -_x}$ ($\alpha=1$). Interestingly, each state's fidelity curve as a function of the valley phase varies in effect size and shape.
Each curve exhibits a wavelike shape, though each has its own amplitude and period. 
Several curves are composed of several frequencies. 
It is outside this paper's scope to explain what causes this variance, focusing instead on the fact that the infidelity amplitude is generally small and can be set to exactly zero by careful valley state preparation.

\begin{figure*}
    \centering
    \begin{subfigure}{0.5\textwidth}
        \centering
        \includegraphics[width=\columnwidth]{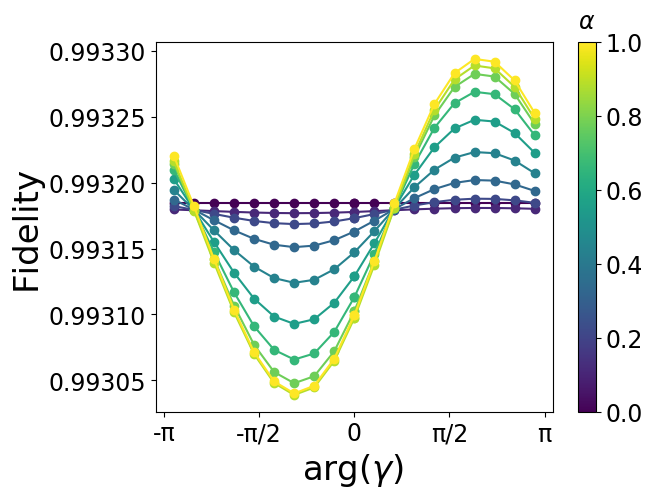}
        \caption{}
    \end{subfigure}%
    ~ 
    \begin{subfigure}{0.5\textwidth}
        \centering
        \includegraphics[width=\columnwidth]{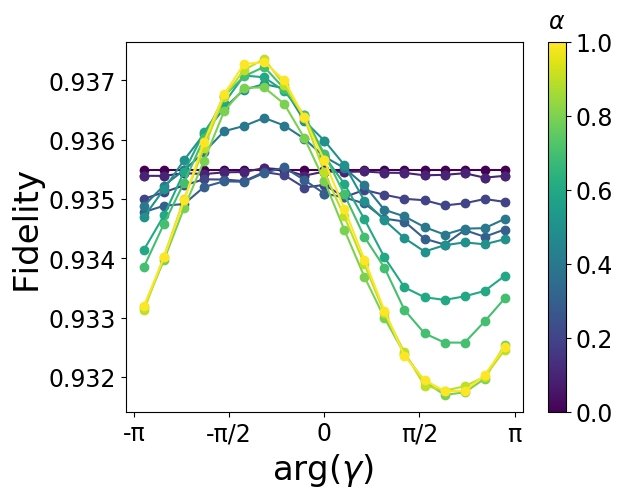}
        \caption{}
    \end{subfigure}
    \caption{
        Average single gate fidelity of a spin SWAP gate across spin chains of (a) $L=20$ and (b) $L=50$. 
        For each plot, the valley $z$ state projection $\alpha$ and valley phase $\arg(\gamma)$ are the variables of interest, and the initial spin state is kept constant.
        Calculations were performed using $|\gamma|=10$, $h=30$, $\Delta=100$ and $J_{\text{SWAP}}=250$.
    }
        \label{fig:SGFidelityAlpha}
\end{figure*}

The crossover between these two behaviors can be investigated by tuning $\alpha$ from 0 to 1, which corresponds to tuning between $\ket{-\ldots -}$ to $\ket{-_x\ldots -_x}$ (Eq.~\ref{eqn:initialstate}). 
The results of this tuning are seen in Fig.~\ref{fig:SGFidelityAlpha}. 
The crossover between these behaviors is smooth, so that if a spin chain suffers from imperfect initialization to a valley $z$ eigenstate, the resulting infidelity will be commensurate with that valley initialization error.

\begin{figure}
    \centering
    \includegraphics[width=0.45\textwidth]{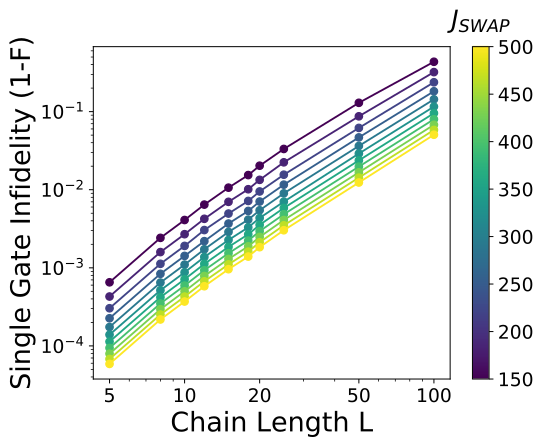}

    \caption{Single gate infidelities ($1-F$) for chains of varying lengths with different values of $J_{\text{SWAP}} \in [150,500]$.
    This simulation was run using $h=750$, $\Delta=500$, $\alpha=0$, and $\gamma=0$, averaging over five random initial spin states. }
    \label{fig:Scaling}
\end{figure}

The effects of electrostatic crosstalk can be isolated by setting $\gamma = 0$.
Crosstalk among the qubits (arising from the electrostatic gate control) represents a significant hurdle for qubit devices, since it is impossible to eliminate crosstalk altogether. 
Residual exchange between unintended dots will always be present, despite the great progress in fidelities made in recent years~\cite{Mills2022,Noiri2022,Takeda2022}.
This crosstalk can be calibrated out experimentally through gate virtualization, but this process is done manually, and to do so for each pair of neighboring dots becomes increasingly impractical as device sizes grow. 
Therefore, understanding the errors that can arise from either miscalibration or the lack of calibration altogether is a current research need. 
In Fig.~\ref{fig:Scaling}, the effects of crosstalk for a wide range of system sizes ($L=[5,100]$) with a uniform residual exchange $J_0$ are shown.
In actuality, this is a generous assumption, as the residual exchange is not usually uniform, and a decreased barrier potential between the two dots for SWAP also affects the exchange coupling between the next nearest neighboring dots. 
Therefore, our calculated fidelities are likely an overestimate.
Again, if the quantitative details underlying crosstalk are available for specific samples, our work can be easily adapted to study specific systems of spin qubits.

It is apparent that single gate fidelities drop to critically low values even for relatively large values of $J_\text{SWAP}$.
This is because the residual exchange $J_0$ crosstalk has more sites to entangle and more time overall to do so.
The compounded errors result in a significant amount of information loss.  
Again, this is the best possible case for electrostatic crosstalk even when valley effects are entirely ignored. 
Reducing the strength of residual exchange relative to the energy scales of intended quantum operations is likely necessary to perform any meaningful quantum circuits on systems larger than ten quantum dots.
This is another important finding of our work

\section{Conclusion}
In conclusion, we investigate the effects of valley leakage on the fidelity of SWAP gates in long quantum spin qubit systems (20-100 qubits).
Using state-of-the-art tensor network methods, the large-scale behavior of our model of a spin qubit chain can be simulated, including both spin, valley, and their coupling as well as crosstalk effects.
We calculate how the fidelity of repeated SWAP gates is directly affected by factors such as valley splitting, spin-valley coupling, Zeeman splitting, and crosstalk. 

Our findings are that the fidelity is only weakly affected by Zeeman splitting and valley splitting except when brought into resonance.
For $z$ basis valley states (or spin states), there is no dependence on the phase of the spin-valley coupling.
When not in the valley $z$ basis due to initialization errors, the effect on fidelity is state-dependent. 
These findings further underscore the importance of reliable valley state initialization.
However, valley phase effects remain a minor correction relative to electrostatic crosstalk.
This is seen in the scaling of the SWAP fidelity as a function of chain length due to the effects of crosstalk alone. 
Based on this scaling behavior, electrostatic crosstalk alone seems a major problem for quantum circuits consisting of more than ten spins, even in the best-case scenario of uniform and constant residual exchange.
Our establishing the tensor network technique as an effective method for studying large spin qubit arrays should have significant future impact on the fabrication and modeling of spin qubits as they scale up to larger sizes.

\section{Acknowledgement}
The authors thank Donovan Buterakos for helpful discussions and suggestions. This work is supported by the Laboratory for Physical Sciences.

\bibliography{mainbib}
\end{document}